# Exact 'antigravity-field' solutions of Einstein's equation

Franklin S. Felber[a]

*Physics Division, Starmark, Inc., P. O. Box 270710, San Diego, California 92198*

Exact time-dependent solutions of Einstein's gravitational field equation for a spherical mass moving with arbitrarily high constant velocity are derived and analyzed. The threshold conditions required for gravitational repulsion of particles at rest are determined as a function of source speed and field strength and particle position.

This paper derives and analyzes exact time-dependent solutions of Einstein's gravitational field equation for a moving mass *in vacuo*. The exact dynamic fields are calculated from an exact metric, which was first derived by Hartle, Thorne, and Price [1], but which has never before been analyzed. This paper shows that the exact 'antigravity' fields calculated here correspond at all source speeds to the weak 'antigravity' fields for any arbitrary velocity and acceleration derived and analyzed by a retarded-potential methodology in [2–4].

In particular, the exact 'antigravity-field' solutions confirm that any mass having a velocity greater than $3^{-1/2}$ times the speed of light gravitationally repels particles at rest within a forward and backward cone, no matter how light the mass or how weak its field.

To calculate the exact strong field of a nonrotating spherically symmetric mass moving with constant relativistic velocity, this paper uses a method for transforming the stationary metric of any mass to the dynamic metric of the mass moving with arbitrarily high constant velocity. Then the geodesic equation is used to calculate the exact strong dynamic field of the mass on a particle at rest. The exact solutions delineate the conditions of source speed and field strength and particle position under which the mass repels particles at rest within a forward and backward cone angle.

The method used for calculating the dynamic gravitational fields of masses moving with arbitrarily high constant velocity is applicable to calculate any field for which the corresponding static or stationary metric of the mass at rest is known. If the *exact* stationary solution of Einstein's field equation is known for a particular mass, then this method gives the *exact* time-dependent solution of Einstein's equation for that mass moving with any constant velocity.

The relativistically exact bound and unbound orbits of test particles in the strong *static* field of a mass at rest have been thoroughly characterized in [5–7], for example. Until recently, calculations [5,6,8,9] of the gravitational fields of arbitrarily moving masses were analyzed only to first order in $\beta$, the ratio of source velocity to the speed of light, $c$. Even in a weak static field, these earlier calculations only solved the geodesic equation for a *nonrelativistic* test particle in the *slow-velocity* limit of source motion. In this slow-velocity limit, the field at a moving test particle has terms that look like the Lorentz field of electromagnetism, called the 'gravimagnetic' or 'gravitomagnetic' field [6,8,9]. Harris [8] derived the *nonrelativistic* equations of motion of a moving test particle in a dynamic field, but only the dynamic field of a *slow-velocity* source. Mashhoon [10] calculated the dynamic gravitomagnetic field of a slowly spinning source having slowly varying angular momentum, and more generally showed explicitly that general relativity contains induction effects at slow source velocities [11].

An exact solution of the field of a relativistic mass is the Kerr solution [5–7,12], which is the exact *stationary* solution for a rotationally symmetric spinning mass. In the stationary Kerr gravitational field, the relativistic unbound orbits of test particles have been approximated in [13].

The first relativistically exact nonstationary field was calculated and analyzed within the weak-field approximation of general relativity in [2]. A Liénard-Wiechert "retarded solution" approach [14] was used in [2] to solve the linearized field equations in the weak-field approximation from the retarded Liénard-Wiechert tensor potential of a relativistic particle. The solution in [2] was used to calculate the weak field acting on a test particle at rest of a mass moving with arbitrary relativistic motion. Since the solution in [2] was the first ever used to analyze the field of a translating mass beyond first order in $\beta$, it was the first to reveal that a mass having a constant velocity greater than $3^{-1/2}c$ gravitationally repels other masses at rest within a cone, as seen by a distant inertial observer. (The Aichelburg-Sexl solution [15] and other boosted solutions, such as in [13], apply only in the proper frame of the accelerating particle, in which no repulsion appears, and not in the laboratory frame.)

This discovery of a repulsive weak gravitational field was used in [3] to calculate the exact relativistic motion of a particle in the strong gravitational field of a mass moving with constant relativistic velocity, but without an explicit calculation of the strong dynamic gravitational field that produced the motion. References [3] and [4] then showed how even a weak repulsive field of a suitable driver mass at relativistic speeds could quickly propel a heavy payload from rest to a speed significantly faster than the driver and close to the speed of light, and do so with manageable stresses on the payload.

In [3], the weak-field dynamic metric on the $x$ axis of a source moving with constant velocity $c\boldsymbol{\beta_0}$ along the $x$ axis was derived from Einstein's equation in the harmonic gauge. In the weak-field approximation, the spacetime interval on the $x$ axis in Cartesian coordinates is [3]

$$ds^2 = [1 + 2(1+\beta_0^2)\Phi]c^2dt^2 - [1 - 2(1+\beta_0^2)\Phi]dx^2 \\ - 8\beta_0\Phi cdtdx - [1 - 2(1-\beta_0^2)\Phi](dy^2 + dz^2) \quad , \quad (1)$$

where the dimensionless potential, $\Phi \equiv -\gamma_0 Gm/(x - \beta_0 ct)c^2$, satisfies the harmonic gauge condition, $\partial\Phi/\partial t + c\beta_0 \partial\Phi/\partial x = 0$; $m$ is the rest mass of the source; and $\gamma_0 \equiv (1-\beta_0^2)^{-1/2}$ is the constant Lorentz factor. The spacetime interval in Eq. (1) is relativistically correct to all orders of $\beta_0$, but is valid only for calculating weak fields and only fields on the $x$ axis.

In the weak-field approximation, in which terms of order $\Phi^2$ are neglected, the geodesic equation applied to the space-time interval of Eq. (1) gives the acceleration of a particle instantaneously at rest on the $x$ axis as [2,3]

$$d^2x/dt^2 \approx -(1-3\beta_0^2)\gamma_0 Gm/(x-\beta_0 ct)^2 , \qquad (2)$$

which is repulsive in both directions, $+x$ and $-x$, for $\beta_0 > 3^{-1/2}$.

---

[a] Electronic mail: felber@san.rr.com

This paper will now use a boost transform method [1] to generalize the spacetime interval of Eq. (1) and the equations of motion: (i) to apply to all space, not just along the axis on which the source moves; (ii) to apply exactly to strong fields as well as weak fields; and (iii) to apply to the gravitational field of any source for which a *distant inertial observer* comoving with the source at speed $\beta_0 c$ in the $+x$ direction would measure a stationary metric, such as the Schwarzschild metric or the Kerr metric. The same method can also be used to find the exact motion of particles following arbitrary relativistic trajectories, not just the motion of particles initially at rest.

In effect, when we, as *distant inertial observers*, measure the static Schwarzschild metric of a mass at rest, we are comoving with the source at constant velocity in the reference frame of any other *distant inertial observer*. Any particle trajectory in the static Schwarzschild field that we observe and record must agree exactly with the same particle trajectory observed and recorded by any other *distant inertial observer*, to within a Lorentz transformation. Recognizing this fact allows exact particle trajectories to be calculated in the strong fields of masses moving with constant relativistic speeds, without having to calculate the strong dynamic fields. Knowing how the physical trajectory in a stationary field appears to a *distant inertial observer* tells immediately how the physical trajectory appears to all other *distant inertial observers*. By this means, particle trajectories in strong dynamic fields were calculated in [3] and [4] and displayed in their accompanying animated solutions.

To generalize Eq. (1) to apply to all space, not just along the $x$ axis on which the source moves, the potential is changed to $\Phi \equiv -\gamma_0^2 Gm/[\gamma_0^2(x-\beta_0 ct)^2 + y^2 + z^2]^{1/2} c^2$. The next step in explicitly calculating the strong dynamic fields is to transform to 'comoving' Cartesian coordinates,

$$\tilde{t} = \gamma_0(ct - \beta_0 x), \quad \tilde{x} = \gamma_0(x - \beta_0 ct), \quad \tilde{y} = y, \quad \tilde{z} = z. \qquad (3)$$

In this comoving coordinate system, the spacetime interval transforms to

$$ds^2 = (1 + 2\Phi/\gamma_0^2)d\tilde{t}^2 - (1 - 2\Phi/\gamma_0^2)(d\tilde{x}^2 + d\tilde{y}^2 + d\tilde{z}^2), \qquad (4)$$

and the potential is $\Phi \equiv -\gamma_0^2 Gm/\tilde{r}c^2$, where $\tilde{r} \equiv (\tilde{x}^2 + \tilde{y}^2 + \tilde{z}^2)^{1/2}$.

The metric in this comoving coordinate system is: (i) diagonal; (ii) independent of $\beta_0$; (iii) stationary, because it is independent of $\tilde{t}$; (iv) static, because it is stationary and time-reversible, that is, invariant under the transformation $\beta_0 \to -\beta_0$; and (v) equal to the weak-field Schwarzschild metric in isotropic coordinates [6].

In isotropic coordinates, the most general exact, static, spherically symmetric solution of Einstein's equation *in vacuo* that reduces to a flat spacetime at large distances ($\tilde{r} \to \infty$) is [6]

$$ds^2 = [(1-\rho)/(1+\rho)]^2 d\tilde{t}^2 - (1+\rho)^4(d\tilde{x}^2 + d\tilde{y}^2 + d\tilde{z}^2), \qquad (5)$$

where $\rho \equiv r_0/\tilde{r}$ and $r_0$ is a constant. The exact spacetime interval of this form reduces in the weak-field approximation to Eq. (4) only for $r_0 = Gm/2c^2$. Perhaps unsurprisingly, the exact field in the comoving coordinate system is just the Schwarzschild field in isotropic coordinates.

Applying the inverse transformation of Eq. (3) to the spacetime interval in Eq. (5) gives the exact dynamic metric of a mass $m$, spherically symmetric in its rest frame, moving with constant velocity $\beta_0 c$ in the $+x$ direction. The only nonvanishing components of the metric in $t, x, y, z$ coordinates are

$$\begin{aligned} g_{00} &= p - \beta_0^2 q, \quad g_{01} = g_{10} = -\beta_0(p-q), \\ g_{11} &= \beta_0^2 p - q, \quad g_{22} = g_{33} = -q/\gamma_0^2, \end{aligned} \qquad (6)$$

where $p \equiv \gamma_0^2[(1-\rho)/(1+\rho)]^2$, and $q \equiv \gamma_0^2(1+\rho)^4$.

The exact dynamic metric of Eq. (6) is a result first derived, but not analyzed, in [1]. The metric has the expected limits, simplifying to: (i) the weak-field metric of Eq. (1) for $\rho \ll 1$; and (ii) the Schwarzschild metric of Eq. (5) for $\beta_0 = 0$. Since the time-dependent metric of Eq. (6) was derived from the exact static metric of Eq. (5) by a coordinate transformation only, Eq. (6) must also exactly satisfy Einstein's equation *in vacuo*, $R_\mu^{\ \nu} - \delta_\mu^{\ \nu} R/2 = 0$, for the Ricci tensor $R_{\mu\nu}$ and scalar curvature $R$.

The event horizon, defined by $g_{00} = 0$ or $1 - \rho = \beta_0(1+\rho)^3$ in these coordinates, is an oblate spheroid compressed by a FitzGerald-Lorentz contraction in the direction of motion, $+x$. A relativistic velocity greatly increases not only the apparent mass of a source, but also, as shown in Fig. 1, the radius of an event horizon $r_h$ in all dimensions.

A simple calculation of an exact strong dynamic gravitational field from Eq. (6) is the field on a particle at rest from a mass $m$ moving along the $x$ axis. Since the 3-velocity of the particle is zero, the exact equations of motion from the geodesic equation are $dt/d\tau = (g_{00})^{-1/2}$ where $\tau$ is the proper time, and, in coordinate time $t$, $d^2x^i/dt^2 + c^2\Gamma^i_{\ 00} = 0$, where $i = 1, 2, 3$, and the $\Gamma^i_{\ 00}$ are Christoffel symbols. The exact strong dynamic gravitational field on a particle at rest at $(t, x, y, z)$, as seen by a *distant inertial observer*, is therefore

$$g_x = \frac{d^2x}{dt^2} = -\left(\frac{1-\rho}{(1+\rho)^7} - \frac{3\beta_0^2}{1-\rho^2}\right)\frac{\gamma_0^4 Gm(x-\beta_0 ct)}{\tilde{r}^3}, \qquad (7a)$$

$$g_y = \frac{d^2y}{dt^2} = -\left(\frac{1-\rho}{(1+\rho)^7} + \frac{\beta_0^2}{1+\rho}\right)\frac{\gamma_0^2 Gmy}{\tilde{r}^3}. \qquad (7b)$$

Equations (7) are an exact solution of Einstein's equation for the field on a particle at rest produced by a mass moving along the $x$ axis. (The $z$ coordinate is equivalent to the $y$ coordinate and is omitted.) In the weak-field approximation ($\rho \approx 0$), Eq. (7a) agrees with [2] and Eq. (2), and the 'antigravity' threshold on the $x$ axis is $\beta_0 \approx 3^{-1/2}$. Equation (7b) confirms the transverse field component is always attractive.

The velocity threshold for 'antigravity' is reduced in strong fields. Figure 2 shows the threshold speed at which the field of the moving mass $m$ reverses on a particle on the $x$ axis, as a function of field strength.

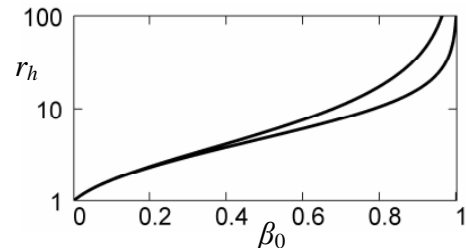

FIG. 1. Event horizon major radius (upper curve) and minor radius (lower curve), normalized to $r_0$, vs. $\beta_0$.

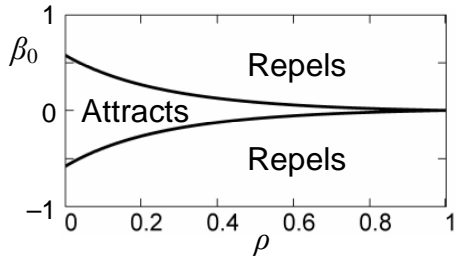 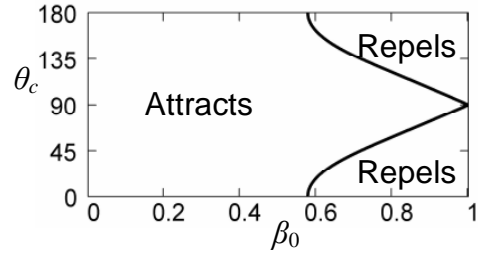

FIG. 2. Threshold speed normalized to *c*, above which particles at rest on *x* axis are repelled, vs. $\rho$.

FIG. 3. Critical half-angle $\theta_c$ (deg), within which the weak radial field of source repels stationary masses, vs. $\beta_0$.

Since the transverse component of the gravitational field is always attractive, a particle at rest will only be repelled by a relativistic mass if the particle lies within a sufficiently narrow angle to the path of a relativistic mass, either in the forward or backward direction. That is, to be repelled, the particle must lie within the surface on which the radial field component $(g_x^2 + g_y^2)^{1/2}$ vanishes. In the weak-field approximation, this critical half-cone angle for the 'antigravity' threshold, measured from the *x* axis is

$$\theta_c = \tan^{-1}[(3\beta_0^2 - 1)^{1/2} / (1 - \beta_0^4)] . \qquad (8)$$

This 'antigravity beam' angle, shown in Fig. 3, is seen to be equivalent to the critical angle calculated in retarded coordinates in [2] by means of the transformation $t = t' + r'/c$, $x - \beta_0 ct = x' - \beta_0 r'$, and $y = y'$, where a prime denotes a retarded quantity (evaluated at the retarded time $t'$), and where $r' = (x'^2 + y'^2)^{1/2}$ is the retarded distance from the source to the observation point, so that $\tilde{r} = \gamma_0(r' - \beta_0 x')$. Expressed in retarded coordinates and in the weak-field approximation, the field components in Eq. (7) become

$$g_x = \frac{-(1 - 3\beta_0^2)\gamma_0 Gm(x' - \beta_0 r')}{(r' - \beta_0 x')^3} , \qquad (9a)$$

$$g_y = \frac{-(1 + \beta_0^2) Gm y'}{\gamma_0 (r' - \beta_0 x')^3} . \qquad (9b)$$

The weak field of a mass with arbitrary velocity, first derived and analyzed in retarded coordinates in [2], is identical, for the special case of constant source velocity along the *x* axis, to the field in Eq. (9). Correspondence is thereby demonstrated between the weak 'antigravity' field of a mass moving with arbitrary velocity in [2–4] and the exact 'antigravity' field in Eq. (7) of a mass moving with constant velocity.

As a simple application of the field derived in Eq. (7), we will now calculate to first order in the field-strength parameter, $\varepsilon \equiv GM/bc^2$, the angular deflection of a particle of mass *m* moving along the *x* axis at nearly constant speed $\beta_0 c$ in the weak static Schwarzschild field of a much larger mass *M*, located at $x = 0$, $y = b$. In the weak-field impulse approximation ($\varepsilon \ll \beta_0^2$) of Eq. (7b), the field on *M* is

$$g_y \approx -(1 + \beta_0^2)\gamma_0^2 Gmb\left[(\gamma_0 \beta_0 ct)^2 + b^2\right]^{-3/2} . \qquad (10)$$

The transverse impulse delivered to the mass *M* is

$$P_y \approx M \int_{-\infty}^{+\infty} g_y dt \approx -2(1 + 1/\beta_0^2)\varepsilon P_0 , \qquad (11)$$

where $P_0 = \gamma_0 mc\beta_0$ is the momentum of the particle. Since $P_y$ is equal and opposite to the impulse delivered to the particle, the angular deflection of the particle is $2(1 + 1/\beta_0^2)\varepsilon$, a result derived in [4] by integrating the orbit equation in a Schwarzschild field. This deflection corresponds to the deflection of a photon in a weak field for $\beta_0 = 1$.

In summary, the first exact time-dependent gravitational-field solutions of Einstein's equation for a moving mass *in vacuo* were derived. This paper presented a general method for transforming any stationary field to a dynamic field that satisfies Einstein's equation. Whether the two-step approach [3,4] to calculating exact orbits in dynamic fields is used, or whether the orbits are calculated directly from the dynamic fields, the exact solutions confirm that even weak gravitational fields of moving masses are repulsive in the forward and backward directions at source speeds greater than $3^{-1/2}c$.

Gravitational repulsion at relativistic speeds opens vistas of opportunities for spacecraft propulsion in the long term [3,4]. Particularly appealing is that propulsion of a massive payload to relativistic speeds can be accomplished quickly and with manageable stresses, because the only stresses in acceleration along a geodesic arise from tidal forces much weaker than the propulsion forces. In the nearer term, exact weak-field dynamic solutions, such as found by the approach outlined here and in [2–4] can be used to test relativistic gravity in the laboratory.

An interesting challenge is to generalize the approach outlined here to calculate the exact dynamic gravitational fields of *accelerating* masses. The exact result should simplify in the weak-field approximation to the fully relativistic weak field of a mass in arbitrary relativistic motion calculated in [2]. Another interesting challenge is to determine the cosmological implications of the new field solutions presented here.

I am grateful to Bahram Mashhoon and Sergei Kopeikin for valuable comments and advice.

1. J. B. Hartle, K. S. Thorne, and R. H. Price in *Black Holes: The Membrane Paradigm*, edited by K.S. Thorne, R. H. Price, D. A. Macdonald (Yale U. Press, New Haven, Conn., 1986), Ch. V.
2. F. S. Felber, http://arxiv.org/abs/gr-qc/0505098 [gr-qc] 2005.
3. F. S. Felber, http://www.arxiv.org/format/gr-qc/0505099, 2005; Space Tech. Applic. Int. Forum – STAIF 2006, M. S. El-Genk, ed., AIP Conf. Proc. **813**, 1374 (12–16 Feb. 2006).
4. F. S. Felber, http://www.arxiv.org/abs/gr-qc/0604076, 2006; Proc. 25th Int. Space Dev. Conf. – ISDC2006 (4–7 May 2006), http://isdc.nss.org/2006/index.html.
5. C.W. Misner, K.S. Thorne, and J.A. Wheeler, *Gravitation* (W.H. Freeman & Co., San Francisco, 1973).
6. H. Ohanian and R. Ruffini, *Gravitation and Spacetime*, 2nd Ed. (W. W. Norton & Co., NY, 1994).
7. S. Chandrasekhar, *The Mathematical Theory of Black Holes* (Oxford University Press, NY, 1983).
8. E. G. Harris, Am. J. Phys. **59**, 421 (1991).
9. I. Ciufolini and J. A. Wheeler, *Gravitation and Inertia* (Princeton University Press, Princeton, 1995).
10. B. Mashhoon, Class. Quantum Grav. **25**, 085014 (2008).
11. D. Bini *et al.*, http://arxiv.org/abs/0803.0390 [gr-qc] 2008.
12. R. P. Kerr, Phys. Rev. Lett. **11**, 237 (1963); R. H. Boyer and R. W. Lindquist, J. Math. Phys. **8**, 265 (1967).
13. C. Barrabès and P. A. Hogan, Phys. Rev. D **70**, 107502 (2004).
14. S. M. Kopeikin and G. Schäfer, Phys. Rev. D **60**, 124002 (1999).
15. P. C. Aichelburg and R. U. Sexl, Gen. Rel. Grav. **2**, 303 (1971).